\documentclass[conference]{IEEEtran}
\IEEEoverridecommandlockouts
\usepackage{cite}
\usepackage{amsmath,amssymb,amsfonts}
\usepackage{algorithm}
\usepackage{algpseudocode}
\usepackage{graphicx}
\graphicspath{ {./images/} }
\usepackage{textcomp}
\usepackage{xcolor}
\usepackage{comment}
\usepackage{booktabs}
\usepackage{multicol}
\usepackage{multirow}
\usepackage{url}
\usepackage{pifont}
\usepackage{orcidlink}
\usepackage{xcolor}
\def\BibTeX{{\rm B\kern-.05em{\sc i\kern-.025em b}\kern-.08em
    T\kern-.1667em\lower.7ex\hbox{E}\kern-.125emX}}
\begin{document}

\title{Deep Ensembling with Multimodal Image Fusion for Efficient Classification of Lung Cancer\\
{
}
\thanks{This research was supported by the J. C. Bose National Fellowship, grant no. JCB/2020/000033 of S. Mitra. (Corresponding author: Surochita Pal.)}
}

\author{\IEEEauthorblockN{Surochita Pal\IEEEauthorrefmark{1} \orcidlink{0000-0002-6749-4426}}
\IEEEauthorblockA{\textit{Machine Intelligence Unit} \\
\textit{Indian Statistical Institute}\\
Kolkata 700108, India \\
Email: pal.surochita@gmail.com}
\and
\IEEEauthorblockN{Sushmita Mitra \orcidlink{0000-0001-9285-1117}}
\IEEEauthorblockA{\textit{Machine Intelligence Unit} \\
\textit{Indian Statistical Institute}\\
Kolkata 700108, India \\
Email: sushmita@isical.ac.in}
}

\maketitle

\begin{abstract}
This study focuses on the classification of cancerous and healthy slices from multimodal lung images. The data used in the research comprises Computed Tomography (CT) and Positron Emission Tomography (PET) images. The proposed strategy achieves the fusion of PET and CT images by utilizing Principal Component Analysis (PCA) and an Autoencoder. Subsequently, a new ensemble-based classifier developed, Deep Ensembled Multimodal Fusion (DEMF), employing majority voting to classify the sample images under examination. Gradient-weighted Class Activation Mapping (Grad-CAM) employed to visualize the classification accuracy of cancer-affected images. Given the limited sample size, a random image augmentation strategy employed during the training phase. The DEMF network helps mitigate the challenges of scarce data in computer-aided medical image analysis. The proposed network compared with state-of-the-art networks across three publicly available datasets. The network outperforms others based on the metrics - Accuracy, F1-Score, Precision, and Recall. The investigation results highlight the effectiveness of the proposed network.
\end{abstract}


\section{Introduction}

For improved treatment options and survival rates early detection of lung cancer is vital. CT scans \cite{jacobson2018computed} provide detailed structural imaging, essential for identifying and diagnosing lung abnormalities. PET scans \cite{dall2021baseline}, on the other hand, offer crucial metabolic data essential for tracking cancer progression and planning treatments. Combining both imaging techniques enhances diagnostic accuracy and improves treatment effectiveness \cite{ambrosini2012pet}. Medical image analysis benefits from multimodal fusion, enhancing precise diagnosis, treatment planning, and improved model performance. Integrating MRI, CT, and PET data optimizes feature extraction, interpretability, and robustness, revealing hidden patterns and reducing noise impact. Fusion strengthens discrimination capabilities, capturing complex relationships for comprehensive data representation.

Deep learning, with automated feature learning capability, has transformed medical image analysis, advancing classification, detection, and diagnosis \cite{pal2024collective}. Unlike traditional methods, deep learning reduces manual feature engineering, accelerates analysis, and enhances precision. Deep learning excels with large datasets, capturing subtle nuances for precise diagnostics. Deep learning standardizes assessments, reducing clinician variability and enabling rapid, accurate diagnoses. Popular deep models like convolutional networks (CNNs), including ResNet, VGG-Net, Inception-Net, and Xception-Net \cite{lecun2015deep, he2016deep, simonyan2015very, szegedy2015going, chollet2017xception} are widely used for automated image analysis tasks. Efficient-Net offers improved classification with fewer parameters \cite{koonce2021efficientnet}. Ensembling, with a diverse range of models, enhances stability and delivers state-of-the-art performance, making ensembling a highly robust approach for a wide array of applications. Deep ensembled classifiers were employed for classification between lung cancer and healthy slices from CT images \cite{shakeel2022automatic}. Techniques like Grad-CAM help visualize influential image regions \cite{selvaraju2017grad} towards earlier disease diagnosis and improved prognosis.

A gradient structural similarity-based filter was devised for multimodal image analysis using CT/MRI data \cite{fu2020gradient}. Analyzing tumor regions extraction in lung images, five optimization algorithms were assessed within the PET-CT framework using CNN \cite{kumar2019co}. Abnormal PET uptake in lungs was found to be significant for tumor detection, highlighting the value of multimodal representations from PET and CT scans, combining anatomical and functional information for more precise diagnosis with reduced false positives/negatives \cite{mercieca2018comparison}. A fully automated pipeline for detecting and segmenting non-small cell lung cancer (NSCLC) was validated using thoracic CT scans from multiple institutions \cite{primakov2022automated}.  

This article introduces a novel Deep Ensembled Multimodal Fusion (DEMF) network, utilizing a sequence of Principal Component Analysis (PCA) and an Autoencoder, referred to as PCAE, for integrating CT and PET scans. PCA retains essential features from each modality, filtering out irrelevant details and optimizing computational resources. The Autoencoder reconstructs fused representations, capturing meaningful patterns efficiently. Overall, PCAE provides a compact, interpretable representation by transforming data into uncorrelated principal components, thereby enhancing efficiency and generalization across diverse applications. The DEMF demonstrates robust generalization even with limited data availability. Main contributions are highlighted below.

\begin{itemize}
    \item Incorporating the PET-CT fusion mechanism PCAE to efficiently extract pertinent reduced features from PET and CT imagery of lung cancer. \begin{itemize} \item The PCA individually extracts the important components of CT and PET modalities, as determined by the PCA graph. The corresponding images are generated in the reduced domains. \item The reconstructed CT and PET images are integrated by the autoencoder to generate the corresponding {\it fused CT-PET} image.
    \end{itemize} 
    \item Creation of an ensemble of foundational classifiers, using majority voting to integrate diverse decisions from the base classifiers. The approach accommodates variability in decisions, stemming from individual classifiers.
    \item Streamlined visualization of activation regions within the framework of multimodal ensembling.
    \item Ablation and comparative analyses showcases the efficacy of the new deep ensembled multimodal classification model, on publicly accessible dataset.  
\end{itemize}
Section \ref{Methodology} describes the fusion technique with ensembled classification in the DEMF network. Section \ref{sec:experiment} provides the experimental implementation and results. Finally Section \ref{sec:discussion} concludes the article. 

\begin{figure*}[!ht]
    \centering
    \includegraphics[width=18cm]{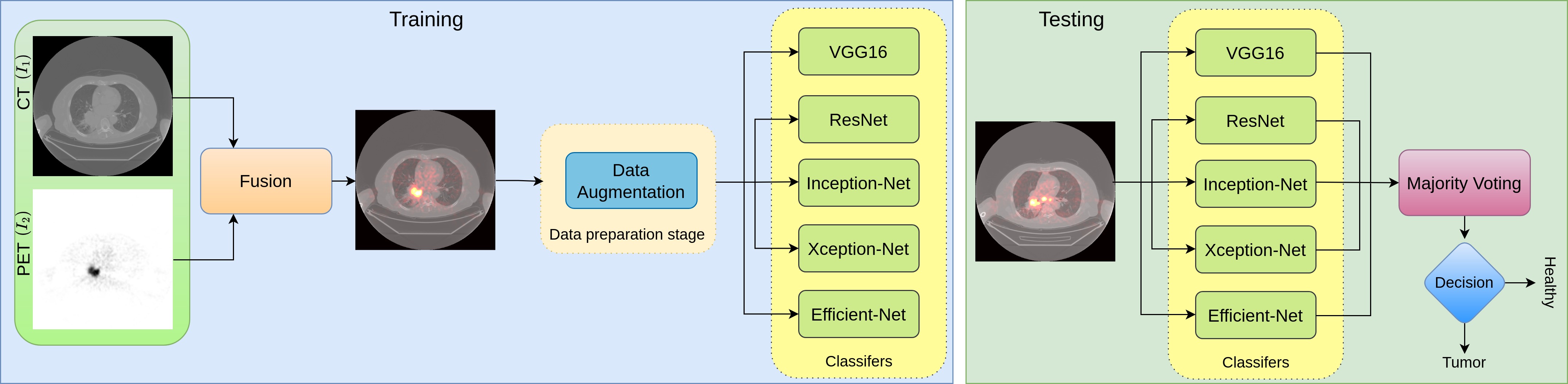}
    \caption{Workflow of ensembled classification, with multimodal image fusion, in DEMF}
    \label{fig:workflowClassifier}
\end{figure*}

\section{Methodology} \label{Methodology}

Remove the pronouns from here "Here we describe the different components of the DEMF network. Initially the CT and PET image features are reduced and fused. Then the integrated image data is augmented (Fig. \ref{fig:augmentedData}), followed by classification through ensembling. The overview of the entire procedure is illustrated in Fig. \ref{fig:workflowClassifier}.

\subsection{Fusion of PET-CT modalities}\label{sec:fusion}

\begin{figure*}[!ht]
    \centering
    \includegraphics[width=18cm]{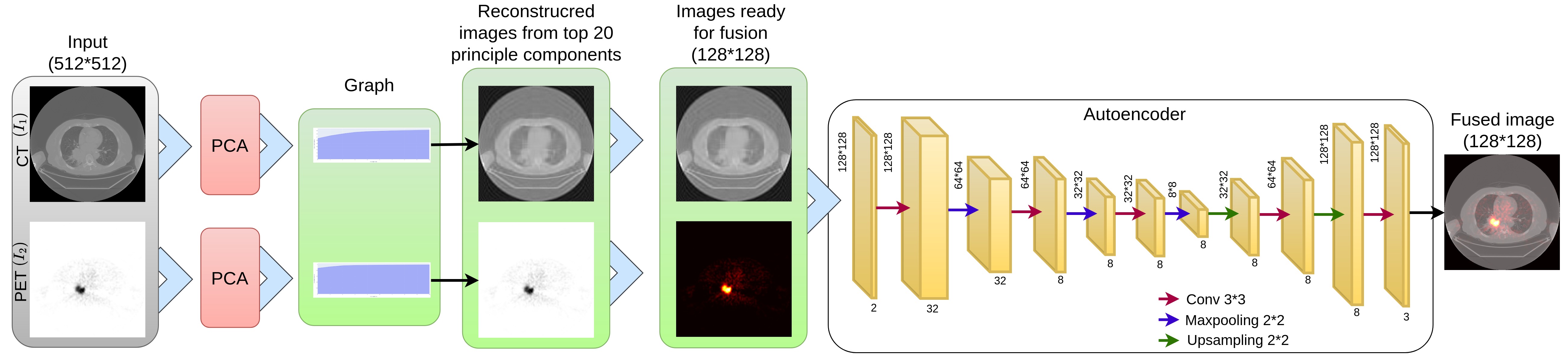}
    \caption{Image fusion strategy}
    \label{fig:fusion}
\end{figure*}

Image fusion in PCAE is accomplished through Principal Component Analysis (PCA), followed by an autoencoder. The method leverages the merits of PCA for dimensionality reduction and feature extraction, along with the capabilities of an autoencoder to discover meaningful patterns from both modalities for efficiently reconstructing the fused representation. The process involves applying PCA to each modality for selecting principal components and then using an autoencoder to learn the relevant patterns, followed by reconstruction of the fused image. A schematic representation of the fusion of CT and PET image modalities is depicted in Fig. \ref{fig:fusion}. 

Consider the two input images, $I_1$ and $I_2$, represented as pixel matrices. These images are integrated using PCA, followed by the autoencoder. Dimensionality reduction is performed on the top 20 principal components, which retain significant relevant information. Then, the pair of $128\times128$ CT and PET images, reconstructed from the principal components, are supplied at the input of the autoencoder shown in Fig. \ref{fig:fusion}. The encoder path consists of three sets of convolution and maxpooling, while the decoder path constitutes two sets of upsampling and convolution. The final outcome of the autoencoder is the decoded {\it fused} image extracted from (20+20=) 40 components.

\subsection{Ensembled classification}\label{sec:CNNarchitecture}

Ensembled deep models combine multiple deep learning algorithms to enhance predictive accuracy, improve generalization, and reduce overfitting. Diverse perspectives foster robust predictions and mitigate individual model biases, resulting in reliable outcomes across complex tasks. The DEMF builds an ensembling of the classifiers VGG16, ResNet, Inception-Net, Xception-Net, and Efficient-Net, pre-trained over the ImageNet, with the final outcome decided by majority voting. Each of these classifiers takes the augmented PET-CT fused image set as input to classify between the cancerous and healthy lung slices. The entire procedure is summarized in Algorithm \ref{alg:ensemble}.

\begin{algorithm}
\caption{Ensembled classification in DEMF}\label{ensemble-majority}
\label{alg:ensemble}

\scalebox{0.8}{
\begin{minipage}{\linewidth}

\small

\begin{algorithmic}[1]

\State $X_i$ = Image $i$
\State $Y_i$ = corresponding label of Image $i$
\State $N$ = Number of images
\State $\text{Data} = \{(X_i, Y_i)\}_{i=1}^{N}$

\#Training

\For{each model in $\text{Ensemble}$}
    \For{each epoch}
        \For{each batch in $\text{Train\_Data}$}
            \State Batch of images $X_b$ and labels $Y_b$
            \State $\text{out} = \text{model}(X_b)$
            \State loss: $\ell = \text{Loss}(Y_b, \text{out})$
            \State gradients: $\nabla_\theta \ell$
            \State weights: $\theta \gets \text{Optimizer}(\theta, \nabla_\theta \ell)$
        \EndFor
    \EndFor
\EndFor

\State Save the trained ensembled model weights

\#Testing
    
\For{each model in $\text{Ensemble}$}
    \State Test\_images $X_v$ and labels $Y_v$
    \State $\text{test\_out} = \text{model}(X_v)$
    \State $\text{test\_loss}$: $\ell_{\text{test}} = \text{Loss}(Y_v, \text{test\_out})$
\EndFor
    
\State $\text{ensemble\_votes}=\text{MajorityVoting}(\text{test\_out})$
\State $\text{acc}_{\text{test}} = \text{Metric}(Y_v, \text{ensemble\_votes})$

\end{algorithmic}
\end{minipage}
}
\end{algorithm}

\normalsize

\section{Implementation and Experimental Results}\label{sec:experiment}

\begin{figure}[!ht]
    \centering
    \includegraphics[width=7cm]{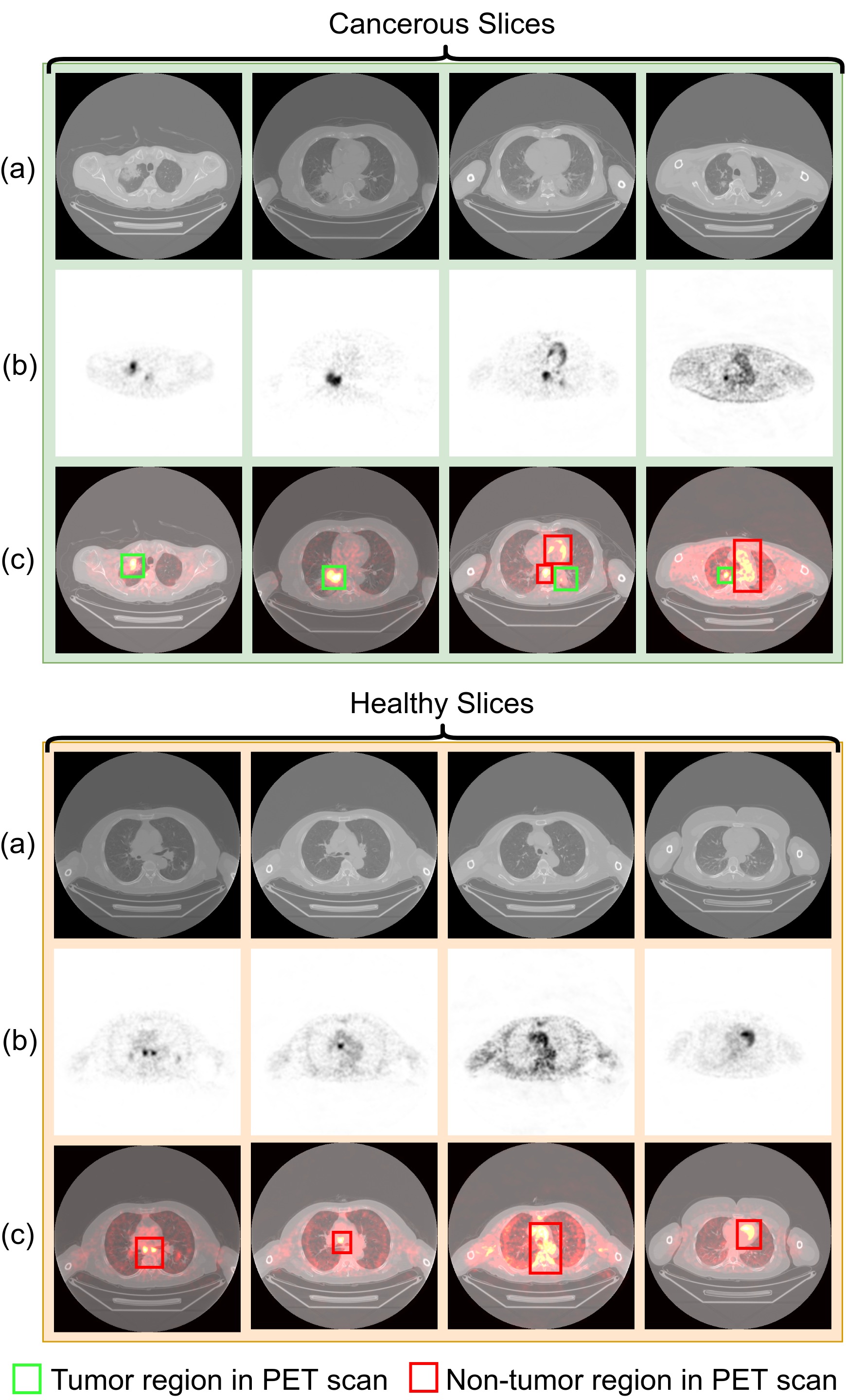}
    \caption{Training samples from (a) CT, (b) PET, and (c) the corresponding fused images, encompassing cancer-affected sample slices along with healthy ones}
    \label{fig:trainingData}
\end{figure}

Three datasets were used in the study. The Figshare data \cite{dong2020improved} comprises seven full-body PET and CT volume scans. The Y-PET/CT \cite{van2021Y90} consists of four lung PET and CT volume data. The Lung-PET-CT-Dx \cite{li2020large} contains 356 CT and PET-CT volumes of the lung, where 133 volumes provide both PET and CT data. All axial slices within the lung region were first extracted from the Figshare data. Next, the multimodal fusion methodology of Section \ref{sec:fusion} was employed. A subset of such combined images, used for training, is shown in Fig. \ref{fig:trainingData}.
The PET scan identifies both tumor and certain non-tumor regions in the lung, with equal intensity values, based on metabolic activity. This makes it challenging to manually distinguish between cancer-affected and healthy tissues. When the effect of both CT and PET is combined, the green bounding boxes (in the figure) become indicative of the actual tumor regions, while the red bounding boxes correspond to the incorrectly highlighted regions which are not essentially tumors. The goal is to locate the tumor region(s) in each of the fused images through the ensembled DEMF.

\begin{figure*}[!ht]
    \centering
    \includegraphics[width=15cm]{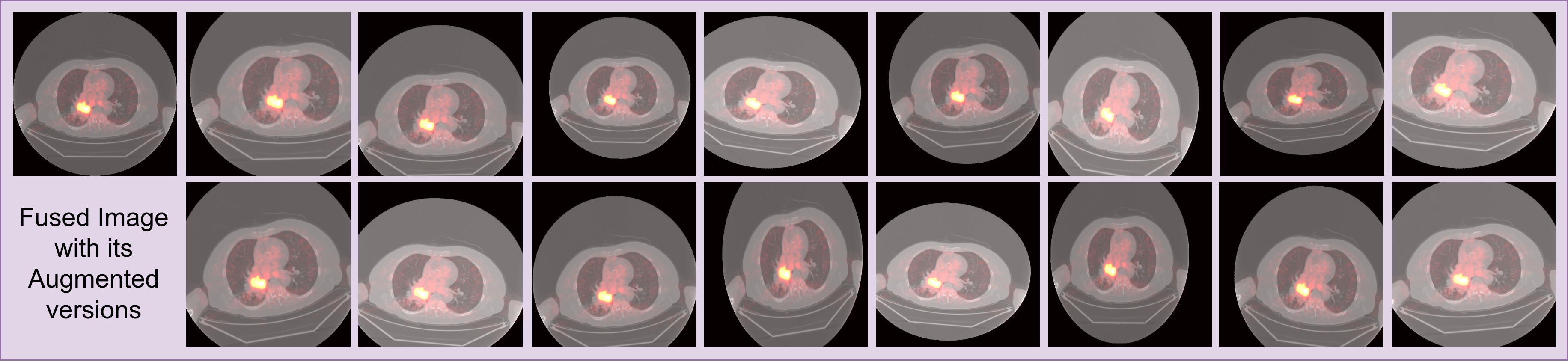}
    \caption{Sample augmentation of fused training data}
    \label{fig:augmentedData}
\end{figure*}

Table \ref{tab:dataSplit} depicts the data distribution for training and testing, with the distribution of cancerous and healthy slices used. The data was split in a 4:1 proportion for training and testing. A 17-fold augmentation was employed for the cancerous slices, as shown in Fig. \ref{fig:augmentedData}, with a 3-fold augmentation being used for the healthy ones. The learning rate was set to $1\times10^{-5}$with a decay rate of $10^{-8}$. The augmentation procedure involved rotation $\pm$ 10, width shift $\pm$ 0.1, height shift $\pm$ 0.1, zoom $\pm$ 0.2, shear $\pm$ 0.2. Any strategy was randomly chosen at each fold.

\begin{table}[!ht]
\caption{Train-test split of the cancerous \& healthy slices of the lung}
\centering
\scalebox{0.80}{
\begin{tabular}{@{}llrcc@{}}
   \toprule \toprule
    Dataset & Purpose &  & Cancerous slices &  Healthy slices \\[0.5ex]
    \midrule
    \multirow{4}{*}{\begin{tabular}{@{}c@{}} Figshare \\ Data \end{tabular}} & & Total $\Rightarrow$ & 80 & 453 \\
    \cline{2-5}
    & \multirow{2}{*}{Training data} & Before Augmentation $\Rightarrow$ & 64 & 362 \\ 
    & & After Augmentation $\Rightarrow$ & 1088 & 1086 \\ \cline{2-5}
    & Testing data & $\Rightarrow$ & 16 & 91 \\ \midrule

    Y-PET/CT & Testing data & $\Rightarrow$ & 43 & 737 \\ \midrule

    \begin{tabular}{@{}c@{}} Lung-PET- \\ CT-Dx \end{tabular} & Testing data & $\Rightarrow$ & 1659 & 6893\\ \midrule
    
    \bottomrule
    \bottomrule
\end{tabular}
}
\label{tab:dataSplit}
\end{table}

The Hounsfield Unit (HU) range was initially maintained in [-1024, 3071] when the pixel values were first normalised. The models were implemented in the Tensorflow framework with a dedicated GPU (NVIDIA TESLA V100 having capacity of 16GB), running behind the wrapper library Keras with Python 3.6.8, Keras 2.2.4, and Tensorflow-GPU version 1.13.1.

\subsection{Classification}

The binary cross-entropy was used, involving batch size of 16 with Adam optimizer. Each individual pre-trained model was further trained upto 100 epochs. 
The metrics used to evaluate the performance of the classifier were Accuracy, F1 score, Precision and Recall. Assuming TP = true positive, TN = true negative, FP = false positive, FN = false negative, we have 
\begin{equation}
\label{metric:Accuracy}
\mathbf{Accuracy} = \biggl(\frac{TP+TN}{TP+TN+FP+FN}\biggr),
\end{equation}

\begin{equation}
\label{metric:F1score}
\mathbf{F1\; score} = \biggl(\frac{2TP}{2TP+(FP+FN)}\biggr), 
\end{equation}

\begin{equation}
\label{metric:Precision}
\mathbf{Precision} = \biggl(\frac{TP}{TP+FP}\biggr),
\end{equation}

\begin{equation}
\label{metric:Recall}
\mathbf{Recall} = \biggl(\frac{TP}{TP+FN}\biggr).
\end{equation}

Table \ref{tab:classificationAccuracy} demonstrates a comparative study of these metrics for evaluating the effectiveness of the ensembled classifier w.r.t. the individual classifier models. It is evident that the ensembled classifer outperformed all the individual classifiers. 

\begin{table}[!ht]
\caption{Comparative study of the individual classifiers and their ensembling, for the fused multimodal image data}
\centering
\scalebox{0.8}{
\begin{tabular}{@{}llcccc@{}}
   \toprule
    \multirow{2}{*}{Dataset} & \multirow{2}{*}{Classifier} & \multicolumn{4}{c@{}}{Metrics} \\ \cline{3-6}
    & & Accuracy \ref{metric:Accuracy} & F1-Score \ref{metric:F1score} & Precision \ref{metric:Precision} & Recall \ref{metric:Recall} \\ \midrule

    \parbox[t]{5mm}{\multirow{6}{*}{\rotatebox[origin=c]{90}{ \begin{tabular}{@{}c@{}} Figshare \\ Data \end{tabular}}}} & VGG 16 & 0.4766 & 0.0968 & 0.0652 & 0.1875 \\
    & ResNet & 0.5234 & 0.1905 & 0.1277 & 0.3750 \\
    & Inception-Net V3 & 0.7850 & 0.3784 & 0.3334 & 0.4375 \\
    & Xception-Net & 0.6262 & 0.2857 & 0.2012 & 0.5003 \\
    & Efficient-Net & 0.8505 & 0.5789 & 0.5070 & 0.6875 \\
    & \textbf{Ensembled DEMF} & \textbf{0.9813} & \textbf{0.9412} & \textbf{0.8889} & \textbf{1.0000} \\
    \midrule

    \parbox[t]{5mm}{\multirow{6}{*}{\rotatebox[origin=c]{90}{ Y-PET/CT }}} & VGG 16 & 0.8269 & 0.3216& 0.2051 & 0.7442  \\
    & ResNet & 0.7163 & 0.1905 & 0.1126 & 0.6190 \\
    & Inception-Net V3 & 0.9436 & 0.6000 & 0.4925 & 0.7674  \\
    & Xception-Net & 0.9154 & 0.5417 & 0.3861 & 0.9070  \\
    & Efficient-Net & 0.9244 & 0.4957 & 0.3919 & 0.6744  \\
    & \textbf{Ensembled DEMF} & \textbf{0.9923} & \textbf{0.9333} & \textbf{0.8936} & \textbf{0.9767} \\
    \midrule

    \parbox[t]{5mm}{\multirow{6}{*}{\rotatebox[origin=c]{90}{ \begin{tabular}{@{}c@{}} Lung-PET- \\ CT-Dx \end{tabular} }}} & VGG 16 & 0.8239 & 0.5786 & 0.5399 & 0.6233 \\
    & ResNet & 0.8313 & 0.6190 & 0.5508 & 0.7064 \\
    & Inception-Net V3 & 0.8214 & 0.6114 & 0.5291 & 0.7239 \\
    & Xception-Net & 0.8736 & 0.6954 & 0.6529 & 0.7438 \\
    & Efficient-Net & 0.9107 & 0.7877 & 0.7308 & 0.8541 \\
    & \textbf{Ensembled DEMF} & \textbf{0.9760} & \textbf{0.9410} & \textbf{0.9003} & \textbf{0.9855}  \\
    \bottomrule

\end{tabular}
}
\label{tab:classificationAccuracy}
\end{table}

\subsection{Qualitative analysis}\label{sec:QualitativeResult}

\begin{figure}[!ht]
    \centering
    \includegraphics[width=8cm]{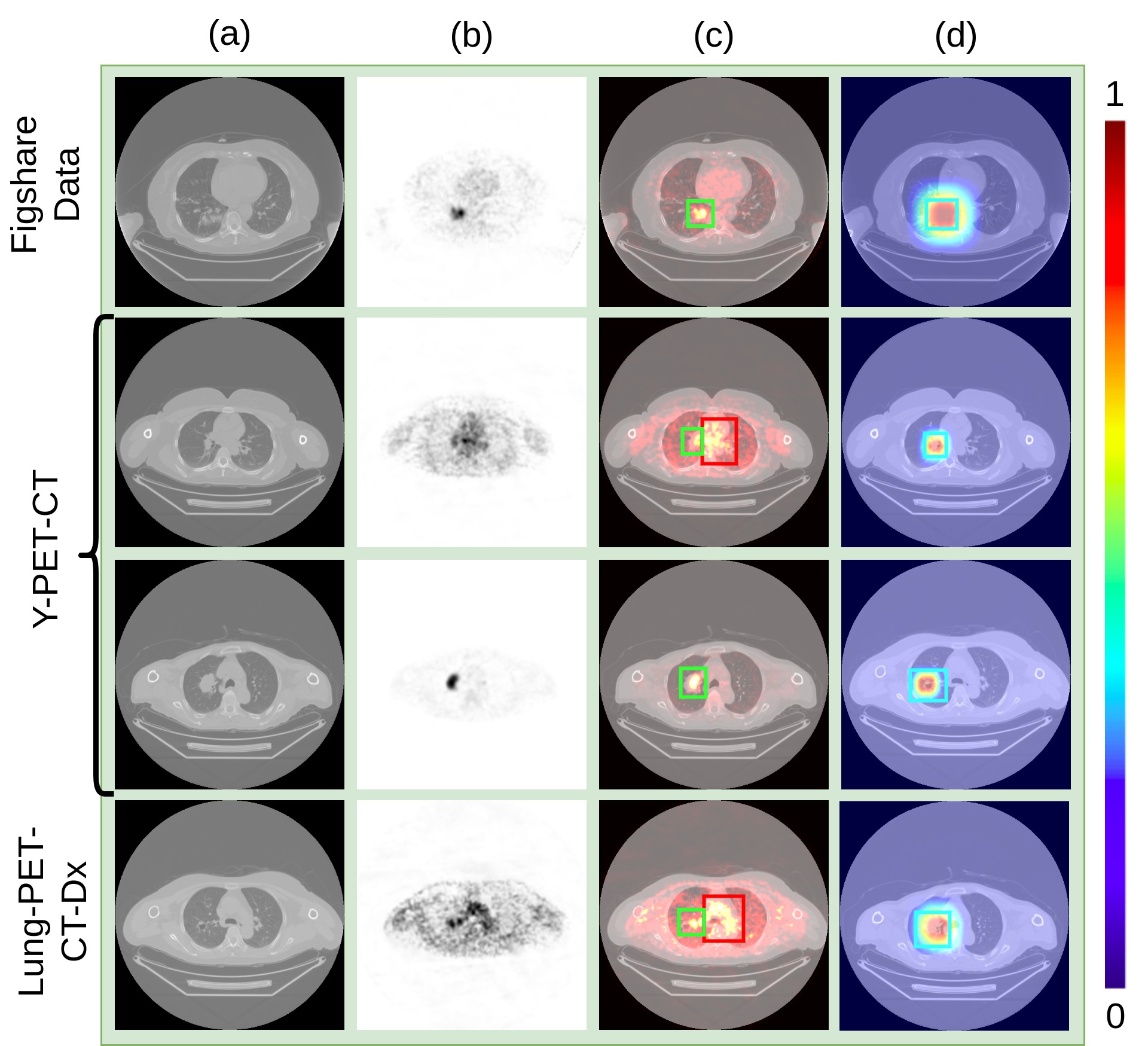}
    \caption{Visualization with Grad-CAM, over sample test set images from (a) CT, (b) PET, and (c) Fused versions, with the corresponding (d) attention map generated by DEMF}
    \label{fig:testingData}
\end{figure}

Fig \ref{fig:testingData} depicts sample visual attention, as generated through Grad-CAM, from the ensembled classifier output over the test dataset. The results demonstrates the robustness of our model performance, in terms of  accurately detecting the cancerous slices while visualizing the exact location of tumor region. The exact location of tumors are depicted by the green bounding box in the fused image, and by the sky-blue bounding box over the correct prediction in the attention map. The red bounding box, in the fused image, is indicative of incorrect detection by PET. This is appropriately corrected in the CT-PET fused image. The importance of the attention map is evident from its precise identification of tumor locations.

\subsection{Ablations} \label{sec:ablation}

\begin{figure}[!ht]
    \centering
    \includegraphics[width=8.5
    cm]{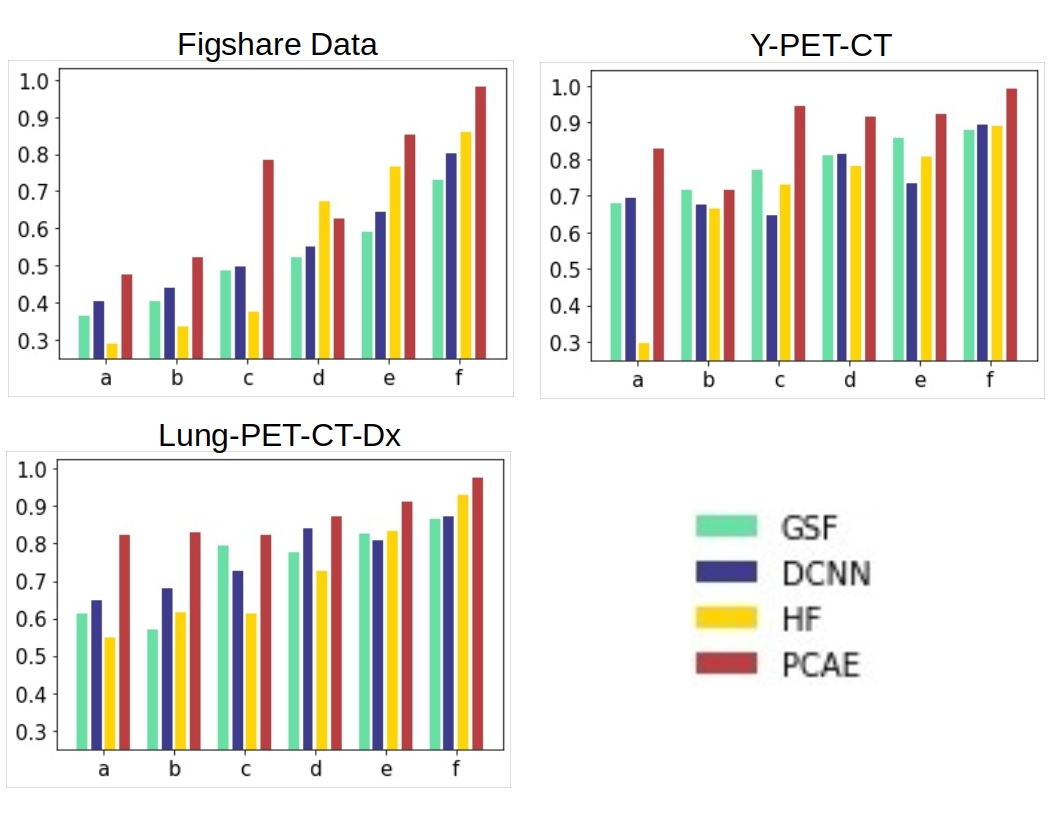}
    \caption{Comparative study of image fusion strategies, in terms of  {\it Accuracy} over each test set, on the performance of (f) ensembled DEMF and the (a) VGG16, (b) ResNet, (c) Inception-Net, (d) Xception-Net, (e) Efficient-Net.}
    \label{fig:ablationComparison}
\end{figure}

Fig. \ref{fig:ablationComparison} illustrates a comparative graphical study of the multimodal fusion strategy PCAE, of ensembled classifier DEMF, with a few other multimodal image fusion techniques over the three test datasets involving CT and PET scans. The fusion strategies compared are GSF \cite{fu2020gradient}, DCNN \cite{xia2019novel}, and HF \cite{xu2020medical}. Fusion  GSF employs structural similarity-based gradient filtering  to emphasize features in source images, for subsequent reconstruction with fused gradients. The DCNN mechanism uses stacked autoencoders to extract the high- and low-frequency components from images, and reintegrates them into the final image. The HF technique uses wavelets to identify optimal parameters of frequencies, for achieving an optimal combination of low- and high-pass filters. The proposed PCAE fusion strategy, combining PCA and Autoencoder, effectively captures input image patterns and features by reducing dimensionality. The PCAE minimizes redundancy, while retaining essential information towards enabling the model to learn hierarchical and abstract features, helping preserve relevant details and offers flexibility across diverse datasets. The comparative performance of these fusion methodologies is evaluated in terms of output $Accuracy$ of individual classifiers like VGG16, ResNet, Inception-Net, Xception-Net, Efficient-Net, as well as the ensembled version DEMF. The fusion result underscores the efficacy of PCAE in improving both image quality and interoperability, thereby leading to enhanced accuracy in classification.  

\begin{figure}[!ht]
    \centering
    \includegraphics[width=7cm]{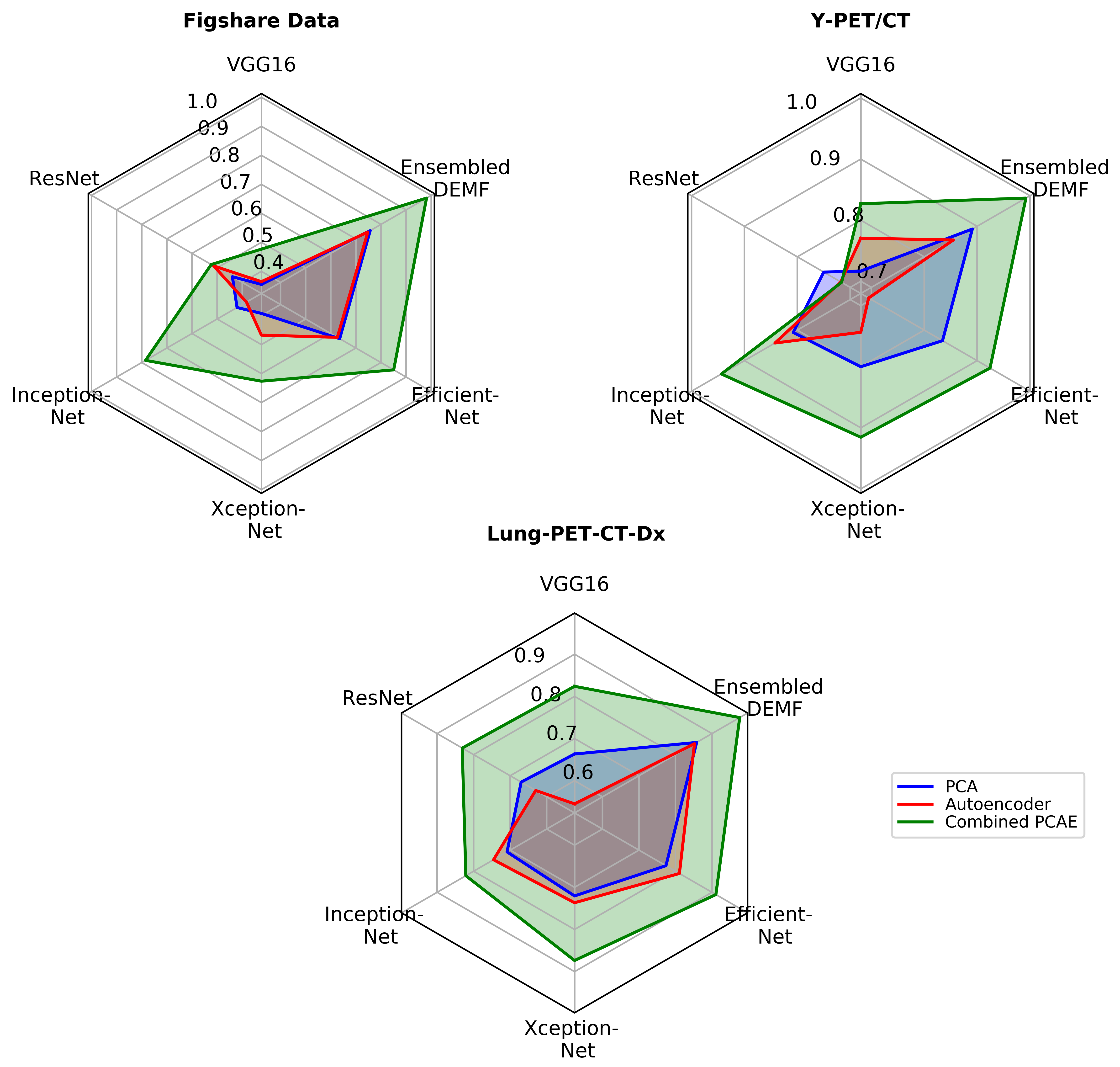}
    \caption{Ablation of multimodal fusion strategies, in terms of {\it Accuracy} by different classifiers}
    \label{fig:ablation1}
\end{figure}

Next the effectiveness of the multimodal image fusion technique was evaluated with respect to each of the individual methodologies, {\it viz.} PCA and Autoencoder. This is  shown in Fig. \ref{fig:ablation1}, in terms of {\it Accuracy} by different classifiers, over all three datasets. It is observed that the proposed fusion methodology always resulted in uniformly good performance, over all the classifiers used.

\begin{figure*}[!ht]
    \centering
    \includegraphics[width=15
    cm]{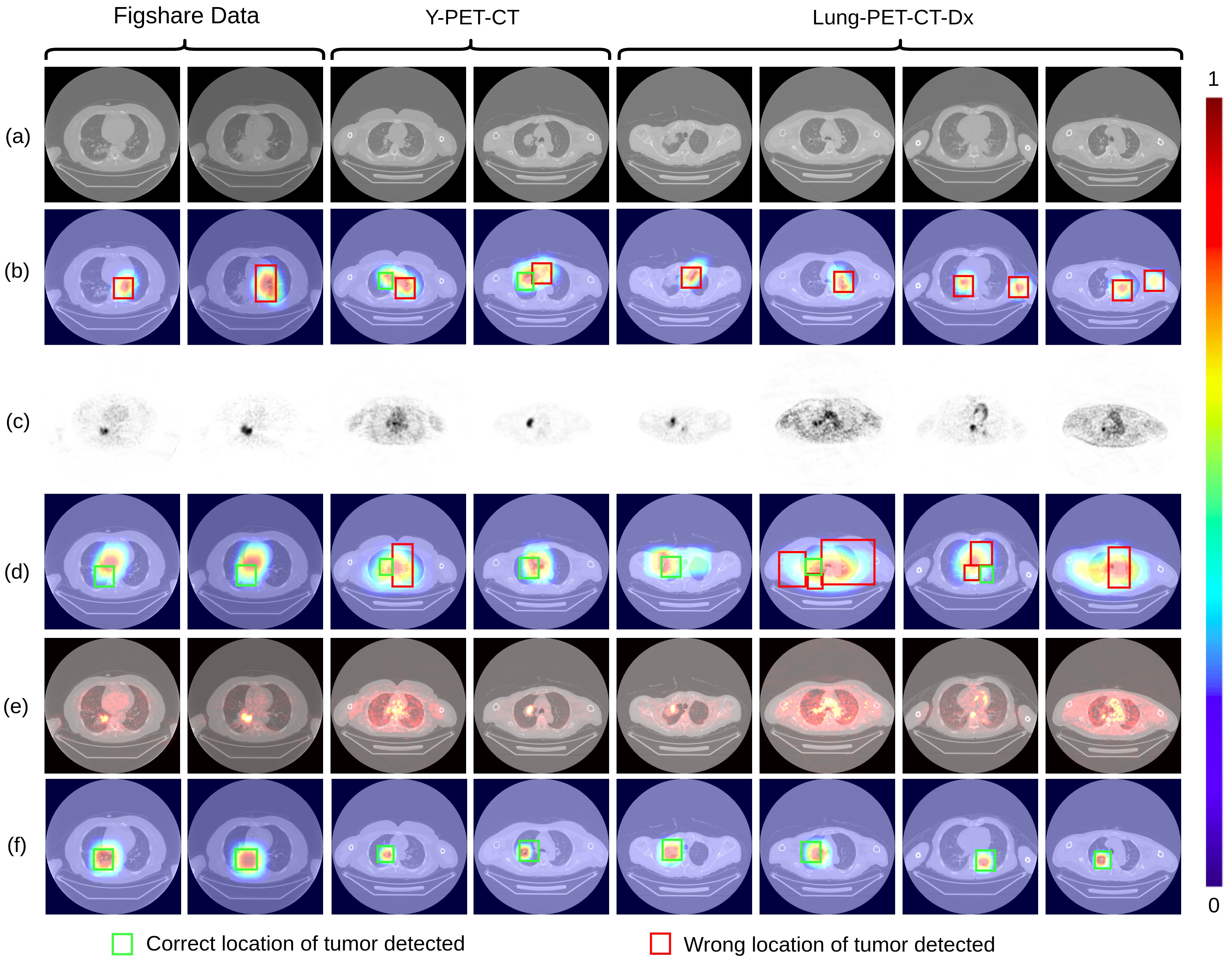}
    \caption{Visualization over sample test cases, using our DEMF with Grad-CAM, in terms of the (a) CT image, with (b) its visualization; the (c) PET image, with (d) its visualization; and the (e) Fused image, with (f) its visualization}
    \label{fig:CTPETablation}
\end{figure*}

The visualization effect on sample test images is depicted in  Fig. \ref{fig:CTPETablation}, over the individual imaging modalities {\it viz.} CT and PET, along with the fused image (incorporating both modalities). Results from the classifier DEMF, trained with only CT images, is shown in  row (b). The observation shows the prediction is often incorrect, as marked by the red bounding boxes. The classifier, trained only over the corresponding PET images, is sometimes found to locate the  correct tumor position in row (d) [green bounding box] and otherwise falsely point to incorrect locations [red bounding box]. On the other hand, using the fused images, the classifier is able to correctly locate the actual tumor region [green box] in row (f). 

\begin{figure}[!ht]
    \centering
    \includegraphics[width=7cm]{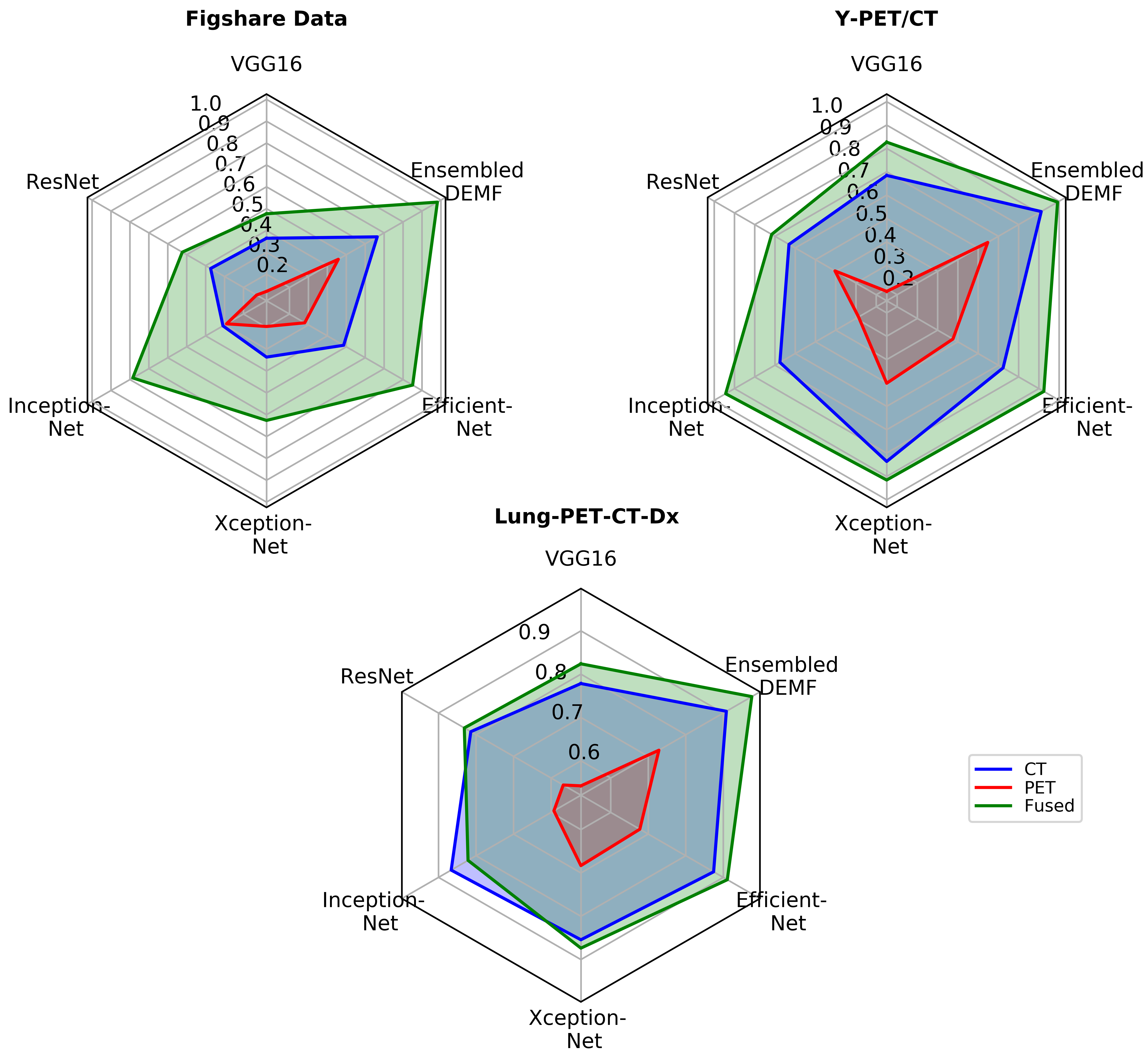}
    \caption{Effect of the imaging modalities, in terms of {\it Accuracy} by the classifiers}
    \label{fig:ablation2}
\end{figure}

An analogous comparative study of the DEMF, with other related deep classifiers, is included in Fig. \ref{fig:ablation2}. Results explores the effect of the individual imaging modalities, along with that of the multimodal fused image, when provided as input to these classifiers. The ensembled classification strategy in DEMF, incorporating the multimodal fusion PCAE, performed the best over all three datasets. 

\section{Conclusions and Discussion}\label{sec:discussion}

A new deep ensembled multimodal fusion network DEMF was developed, for effectively classifying lung CT slices as cancerous or healthy.  The ensembled model combined a number of distinct classifiers using majority voting. 
The approach was robust and less prone to overfitting, as relied on a collective decision from diverse models. Majority voting was computationally efficient and simple to implement. The procedure enhanced generalization by reducing the risk of memorizing noise or outliers in the limited training data. 
Ablation studies and visualization revealed the ensembled deep multimodal classifier helped emphasize cancerous regions within the images.

The fusion strategy PCAE helped reduce dimensionality, involving the PCA and autoencoder components. Leveraging multimodal data with reduced feature cardinality, and ensembled fusion, generated more resilient results. This methodology can be applied to various diseases affecting different organs and can incorporate multiple imaging modalities.

\section*{Declaration} The authors declare no conflict of interest.

\bibliographystyle{IEEEtran}
\bibliography{reference}

\begin{thebibliography}{10}
\providecommand{\url}[1]{#1}
\csname url@samestyle\endcsname
\providecommand{\newblock}{\relax}
\providecommand{\bibinfo}[2]{#2}
\providecommand{\BIBentrySTDinterwordspacing}{\spaceskip=0pt\relax}
\providecommand{\BIBentryALTinterwordstretchfactor}{4}
\providecommand{\BIBentryALTinterwordspacing}{\spaceskip=\fontdimen2\font plus
\BIBentryALTinterwordstretchfactor\fontdimen3\font minus \fontdimen4\font\relax}
\providecommand{\BIBforeignlanguage}[2]{{%
\expandafter\ifx\csname l@#1\endcsname\relax
\typeout{** WARNING: IEEEtran.bst: No hyphenation pattern has been}%
\typeout{** loaded for the language `#1'. Using the pattern for}%
\typeout{** the default language instead.}%
\else
\language=\csname l@#1\endcsname
\fi
#2}}
\providecommand{\BIBdecl}{\relax}
\BIBdecl

\bibitem{jacobson2018computed}
F.~L. Jacobson and M.~T. Jaklitsch, ``Computed tomography scanning for early detection of lung cancer,'' \emph{Annual Review of Medicine}, vol.~69, pp. 235--245, 2018.

\bibitem{dall2021baseline}
F.~G. Dall’Olio, D.~Calabr{\`o} \emph{et~al.}, ``{Baseline total metabolic tumour volume on 2-deoxy-2-[18F] Fluoro-D-Glucose Positron Emission Tomography-Computed Tomography as a promising biomarker in patients with advanced non--small cell lung cancer treated with first-line pembrolizumab},'' \emph{European Journal of Cancer}, vol. 150, pp. 99--107, 2021.

\bibitem{ambrosini2012pet}
V.~Ambrosini, S.~Nicolini \emph{et~al.}, ``{PET/CT} imaging in different types of lung cancer: {A}n overview,'' \emph{European Journal of Radiology}, vol.~81, pp. 988--1001, 2012.

\bibitem{pal2024collective}
S.~Pal, S.~Mitra, and B.~U. Shankar, ``Collective intelligent strategy for improved segmentation of {COVID-19} from {CT},'' \emph{Expert Systems with Applications}, vol. 235, p. 121099, 2024.

\bibitem{lecun2015deep}
Y.~LeCun, Y.~Bengio, and G.~Hinton, ``Deep learning,'' \emph{Nature}, vol. 521, pp. 436--444, 2015.

\bibitem{he2016deep}
K.~He, X.~Zhang \emph{et~al.}, ``Deep residual learning for image recognition,'' in \emph{Proceedings of the IEEE Conference on Computer Vision and Pattern Recognition}, 2016, pp. 770--778.

\bibitem{simonyan2015very}
K.~Simonyan and A.~Zisserman, ``Very deep convolutional networks for large-scale image recognition,'' in \emph{Proceedings of the 3rd International Conference on Learning Representations (ICLR 2015)}, 2015, pp. 1--14.

\bibitem{szegedy2015going}
C.~Szegedy, W.~Liu \emph{et~al.}, ``Going deeper with convolutions,'' in \emph{Proceedings of the IEEE Conference on Computer Vision and Pattern Recognition}, 2015, pp. 1--9.

\bibitem{chollet2017xception}
F.~Chollet, ``{Xception: Deep learning with depthwise separable convolutions},'' in \emph{Proceedings of the IEEE Conference on Computer Vision and Pattern Recognition}, 2017, pp. 1251--1258.

\bibitem{koonce2021efficientnet}
B.~Koonce, \emph{{EfficientNet}}.\hskip 1em plus 0.5em minus 0.4em\relax Apress, 2021, pp. 109--123.

\bibitem{shakeel2022automatic}
P.~M. Shakeel, M.~Burhanuddin \emph{et~al.}, ``{Automatic lung cancer detection from CT image using improved deep neural network and ensemble classifier},'' \emph{Neural Computing and Applications}, vol.~34, pp. 9579–--9592, 2022.

\bibitem{selvaraju2017grad}
R.~R. Selvaraju, M.~Cogswell \emph{et~al.}, ``{Grad-CAM: Visual explanations from deep networks via gradient-based localization},'' in \emph{Proceedings of the IEEE International Conference on Computer Vision}, 2017, pp. 618--626.

\bibitem{fu2020gradient}
Z.~Fu, Y.~Zhao \emph{et~al.}, ``Gradient structural similarity based gradient filtering for multi-modal image fusion,'' \emph{Information Fusion}, vol.~53, pp. 251--268, 2020.

\bibitem{kumar2019co}
A.~Kumar, M.~Fulham \emph{et~al.}, ``Co-learning feature fusion maps from {PET-CT} images of lung cancer,'' \emph{IEEE Transactions on Medical Imaging}, vol.~39, pp. 204--217, 2019.

\bibitem{mercieca2018comparison}
S.~Mercieca, J.~Belderbos \emph{et~al.}, ``{Comparison of $SUVmax$ and $SUVpeak$ based segmentation to determine primary lung tumour volume on FDG PET-CT correlated with pathology data},'' \emph{Radiotherapy and Oncology}, vol. 129, pp. 227--233, 2018.

\bibitem{primakov2022automated}
S.~P. Primakov, A.~Ibrahim \emph{et~al.}, ``Automated detection and segmentation of non-small cell lung cancer computed tomography images,'' \emph{Nature Communications}, vol.~13, p. 3423, 2022.

\bibitem{dong2020improved}
Y.~Dong, W.~Yang \emph{et~al.}, ``{An improved supervoxel 3D region growing method based on PET/CT multimodal data for segmentation and reconstruction of GGNs},'' \emph{Multimedia Tools and Applications}, vol.~79, pp. 2309--2338, 2020.

\bibitem{van2021Y90}
B.~J. Van, Y.~K. Dewaraja \emph{et~al.}, ``{Y-90 SIRT: Evaluation of TCP variation across dosimetric models},'' \emph{EJNMMI Physics}, vol.~8, p.~45, 2021.

\bibitem{li2020large}
P.~Li, S.~Wang \emph{et~al.}, ``{A large-scale CT and PET/CT dataset for lung cancer diagnosis [dataset]},'' \emph{The Cancer Imaging Archive}, 2020.

\bibitem{xia2019novel}
K.~Xia, H.~Yin \emph{et~al.}, ``A novel improved deep convolutional neural network model for medical image fusion,'' \emph{Cluster Computing}, vol.~22, pp. 1515--1527, 2019.

\bibitem{xu2020medical}
L.~Xu, Y.~Si \emph{et~al.}, ``Medical image fusion using a modified shark smell optimization algorithm and hybrid wavelet-homomorphic filter,'' \emph{Biomedical Signal Processing and Control}, vol.~59, p. 101885, 2020.

\end{thebibliography}

\end{document}